\documentclass[%
 reprint,
 amsmath,amssymb,
 aps,
]{revtex4-2}
\usepackage[utf8]{inputenc}
\usepackage{graphicx}
\usepackage{dcolumn}
\usepackage{bm}
\usepackage{hyperref}
\usepackage{amssymb}
\hypersetup{
    colorlinks=true,
    citecolor=blue,
    linkcolor=blue,
    filecolor=magenta,
    urlcolor=blue,
 }

\begin{document}
\preprint{APS/123-QED}
\title{A new dynamical mechanism of incomplete fusion in heavy-ion collision}

\author{A.K. Nasirov$^{1,2}$}
 \email{nasirov@jinr.ru}
\author{B.M. Kayumov$^{2,3}$}
\author{O.K. Ganiev$^{2,4,5}$}
\author{G.A. Yuldasheva$^{2}$}
\affiliation{
$^1$Joint Institute for Nuclear Research, 141980 Dubna, Russia \\
$^2$Institute of Nuclear Physics, Uzbekistan Academy of Sciences,  100214 Tashkent,
Uzbekistan \\
$^3$New Uzbekistan University, 100007 Tashkent, Uzbekistan\\
$^4$School of Engineering, Akfa University, 111221 Tashkent, Uzbekistan\\
$^5$Faculty of Physics, National University of Uzbekistan, 100174 Tashkent, Uzbekistan}

\date{\today}

\begin{abstract}
The incomplete fusion has been proved as  the formation and emission of the $\alpha$ 
particle  by the increase in the rotational energy of the very mass-asymmetric
dinuclear system. 
The results of the dinuclear system model have confirmed that the
incomplete fusion in heavy-ion collisions occurs at a large orbital
angular momentum ($L > 30 \hbar$) due to the strong increase 
of the intrinsic fusion barrier.
\end{abstract}

\maketitle


The incomplete fusion (ICF) of nuclei at the heavy-ion collision is observed
 in reactions of light projectiles with the intermediate-mass target
nucleus. This phenomenon was first observed more than 60 years ago
 \cite{Knox1960}.
 From the analysis of experimental data it was found that the peripheral
 collisions are a favorable condition for the incomplete fusion. This phenomenon
 is studied by observation of the $\alpha$ particle flying in the forward angles
 or other light clusters  or by identification of the evaporation residue accompanied
 with the emitted fast light clusters.

  A mean value $<L> = 40 \hbar$ of the angular momentum distribution of the entrance
channel corresponding to the incomplete fusion was observed
 for the $^{159}$Tb($^{14}$N, $\alpha x$n)$^{169-x}$Yb reaction  \cite{Inamura1977},
 while the estimated value \\ $<L> = 30 \hbar$ was presented in Ref. \cite{Barker}
 for the $^{16}$O+$^{146}$Nd reaction.
 The cross section of the evaporation residues (ER) formed in the incomplete fusion
 increases gradually with the mass asymmetry of the reaction entrance channel
 \cite{Kumar2019}.
 The ER presence accompanied with the emission of the $\alpha$ particle formed in the
 incomplete fusion has been established from the analysis of the total cross section
 for the $\alpha$-particle production
 by the standard statistical models ALICE-91 \cite{Kumar2019} and
 PACE4 \cite{Kumar2013,Agarwal}. The conclusion of the authors
 is based on the enhancing underestimation of the
 measured cross sections of $\alpha$-emitting residues  by the theoretically predicted
 cross sections.

 The breakup fusion model \cite{Udagawa1980}  was suggested to describe ICF.
In this model, the projectile is assumed to break
up into a light cluster and a conjugate nucleus at close distances to the
 target nucleus.

The sum-rule model \cite{Wilczynski1980, Wilczynski1982} developed
  by Wilczynski {\it et al.} was used to calculate
the ER cross section of various projectile-like fragments formed
in ICF reactions. The authors concluded that ICF reactions are localized in the
angular momentum space above the critical angular momentum
 $\ell_{\rm cr}$ for complete fusion (CF) of projectile and target.

  The breakup of the projectile was analysed in the recent paper \cite{Bosshe2019}
  by R. V. den Bosshe and A.D. Torres by combining a classical trajectory model with
  stochastic breakup with the quantum-mechanical fragmentation theory \cite{Kuklin2012}
treatment of two-body clusterization and decay of a projectile.
  The angular distributions of the clusters $^{4}$He and $^{8}$Be produced
  at the breakup of projectile in the $^{20}$Ne+$^{208}$Pb reactions explored
  and compared with the experimental data.

 We should stress that the projectile breakup mechanism of the incomplete fusion
 was assumed in all of the above-listed theoretical methods to analyse measured data.

\begin{figure*}[t!]
\centering
\vspace{-0.7cm}
  \includegraphics[width=0.75\textwidth]{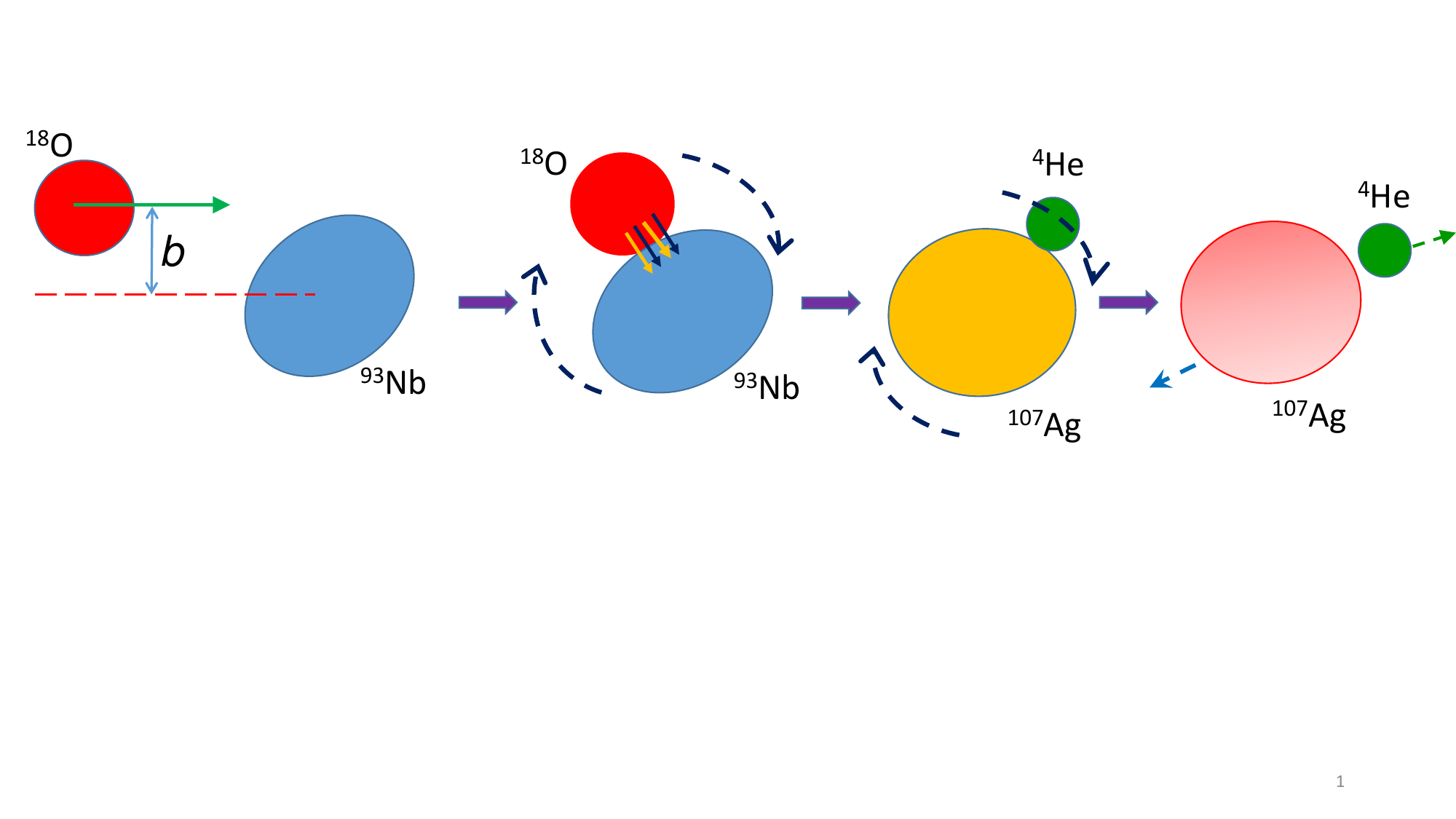}
\vspace{-3.6cm}
\caption{\label{ICF} The sketch of the incomplete fusion mechanism as very
asymmetric quasifission in the case of the $^{18}$O+$^{93}$Nb reaction.
}
\end{figure*}

 In this work, we consider a new mechanism of the ICF reaction
  as a quasifission to  calculate the excitation function of  the evaporation residues
 (ER) survived against fission, which is accompanied by the  $\alpha$-particle emission from the dinuclear system (DNS)  (see Fig. \ref{ICF}).
  The DNS is formed at the capture(full momentum transfer) of the projectile nucleus by the target nucleus.

This mechanism is based on the DNS concept of the complete fusion which operates
 with such physical quantities as intrinsic fusion barrier $B^*_{\rm fus}$,
 quasifission barrier  $B_{\rm qf}$ and the excitation energy $E^*_Z$ of the DNS
 with the charge asymmetry $Z$.
 The DNS evolution by the diffusion process due to the nucleon transfer between
 fragments leads to the formation of the $\alpha$ particle in collisions
  with the large orbital angular momentum if capture takes place at the given beam energy.

\begin{figure}[t!]
\centering
  \includegraphics[width=0.47\textwidth]{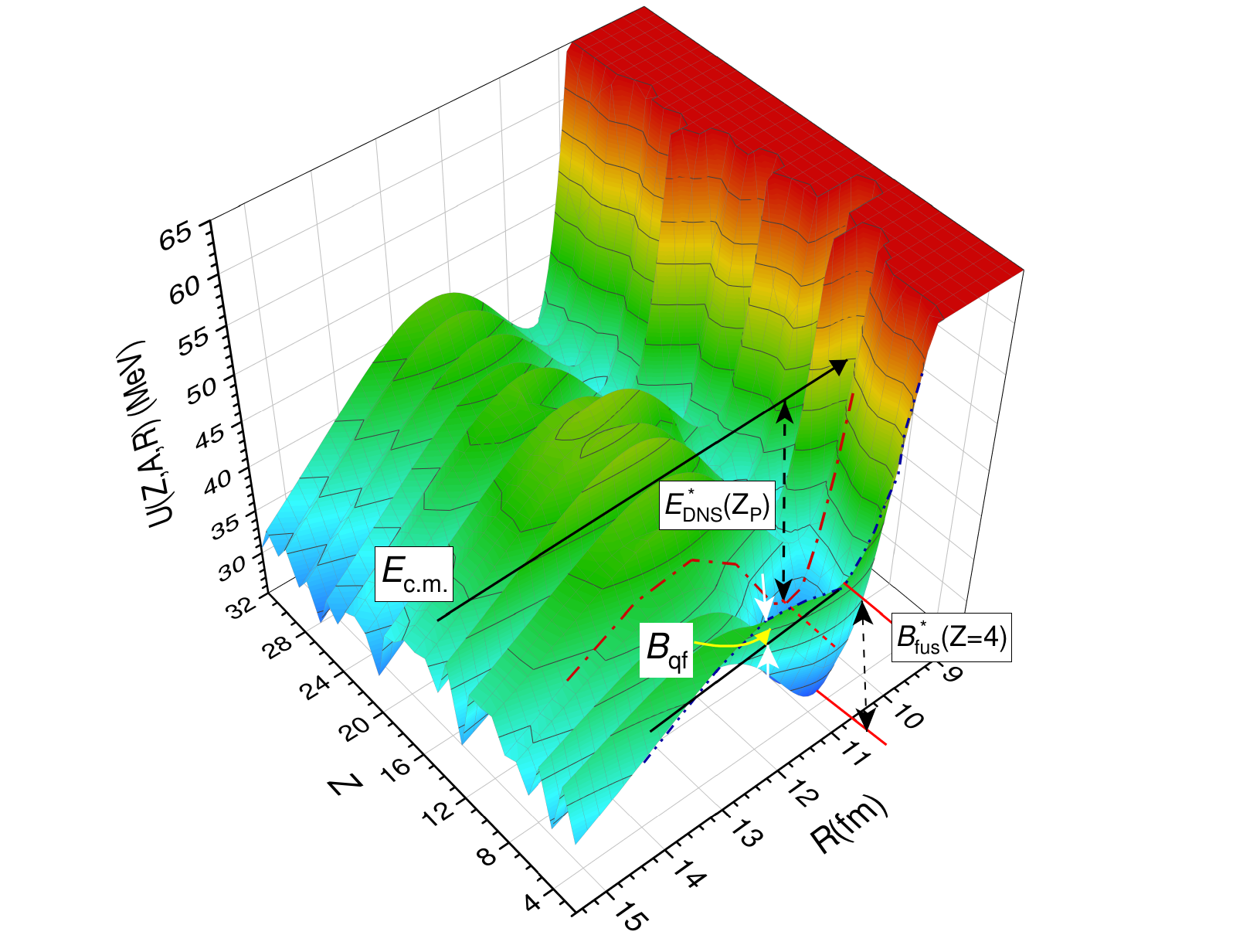}
  \vspace{-0.2cm}
\caption{\label{BqfBfus} Potential energy surface calculated for the
DNS formed in the $^{16}$O+$^{130}$Te reaction at the collisions
 with the values of orbital angular momentum $L=40 \hbar$
 as a function of the fragment charge numbers ($Z$) and relative distance
 ($R$) between centres-of-mass fragments. The DNS excitation energy $E^*_{\rm DNS}(Z_P)$,
 the intrinsic fusion $B^*_{\rm fus}$ barrier is shown for $Z=4$ and
 quasifission $B_{\rm qf}$ barrier is shown for $Z=2$ by the corresponding arrows.}
\end{figure}

\begin{figure}[t!]
\centering
  \includegraphics[width=0.47\textwidth]{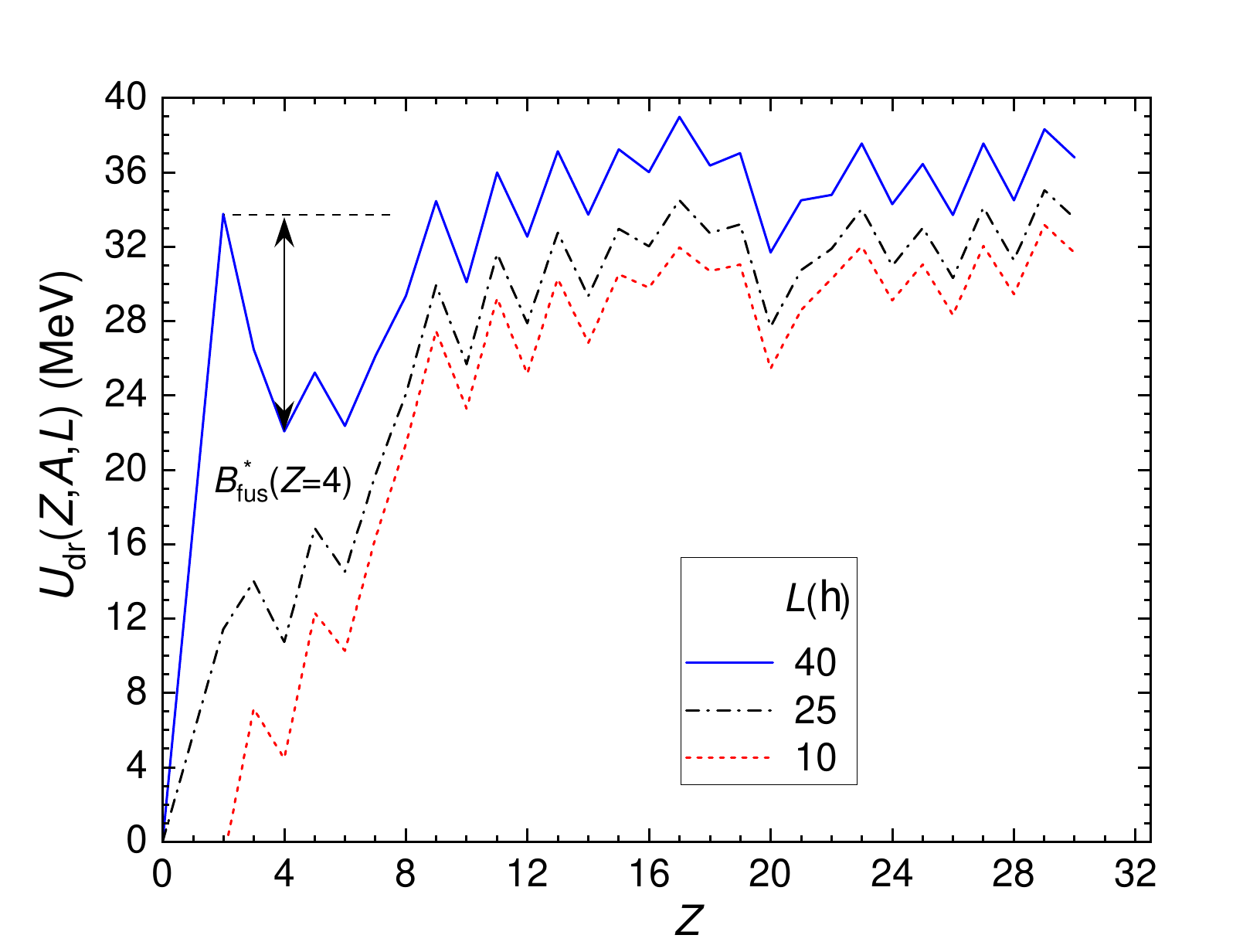}
  \vspace{-0.2cm}
\caption{\label{BfusUdr} The driving potential of the DNS formed
 in the $^{16}$O+$^{130}$Te reaction calculated for the orbital angular momentum
  $L=10, 25, and 40\hbar$.
 the intrinsic fusion $B^*_{\rm fus}$ and quasifission $B_{\rm qf}$ barriers of the
 entrance channel $Z_P$ are shown by the corresponding arrows.}
\end{figure}

  The increase of the orbital angular momentum in the entrance channel leads to the following changes of the physical quantities causing the enhance
  of the incomplete fusion probability.

 i) The increase of the dynamical intrinsic barrier $B^*_{\rm fus}$ to complete fusion at the very asymmetric  charge and mass distribution corresponding to the $\alpha$ particle (see Figs.  \ref{BqfBfus} and \ref{BfusUdr}.

 ii) The decrease of the stability of the DNS due to decrease the depth of the potential well of the nucleus--nucleus interaction. It is called the quasifission barrier
      $B_{\rm qf}$ (see Fig. \ref{BqfBfus}).

 iii) The excitation energy $E^*_Z$ of the DNS with the charge asymmetry $Z$,
       which is generated from the total kinetic energy loss at the capture of
       the projectile by the target nucleus, decreases due to increase
       the DNS rotational energy (see Fig. \ref{BqfBfus}).
        Therefore, the residue nucleus formed in the incomplete fusion is
        less heated than the compound nucleus formed in the
        complete fusion.

  iv)  The effect of the centrifugal force on the very mass-asymmetric DNS enhances
   leading to the incomplete fusion which can be considered as the DNS quasifission  producing very mass-asymmetric products.


  The potential energy surface (PES) presented in Fig. \ref{BqfBfus} is calculated as a
  sum  of the  nucleus--nucleus interaction $V$  and reaction energy balance $Q_{gg}$
  \cite{Kayumov2022}:
\begin{eqnarray}\label{drivpot}
U(Z,A,L,R,\{\beta_i,\alpha_i\})&=&Q_{gg}-V_{\rm rot}^{\rm CN}(L)\nonumber\\
&+&V(Z,A,L,R,\{\alpha_i,\beta_i\})
\end{eqnarray}
  where $Z$ and $A$  are charge and mass numbers, respectively, of
  a DNS fragment, and the ones of the conjugate fragment are
  $Z_c=Z_{\rm CN}-Z$ and $A_c=A_{\rm CN}-A$;
$Q_{gg}=B_1+B_2-B_{\rm CN}$ is the reaction energy
balance; $B_1$, $B_2$, and $B_{CN}$ are the binding energies of the interacting
nuclei and CN, respectively, which are obtained from the nuclear mass tables in Refs.
\cite{Audi1995,Moeller1995};
$\beta_i$ and   $\alpha_i$ represent  deformation parameters
 (quadrupole and octupole) of the DNS fragments and orientation angles
   of the axial symmetry axis of the deformed nuclei to the beam
   direction, respectively.
 In case of the nuclei with at spherical shape at their ground state,
 the change in their shape due to surface vibration at the zero-point motion
 is considered \cite{Kayumov2022}. The amplitudes of vibrations are taken
 equal to the values  of the deformation parameters of the first quadrupole $2^+$ and
 octupole $3^-$ collective excitations of nuclei
 ($\beta^+_2$) ~\cite{Raman}  and  ($\beta^-_3$) \cite{Spear}.

\begin{figure}[t!]
\centering
  \includegraphics[width=0.47\textwidth]{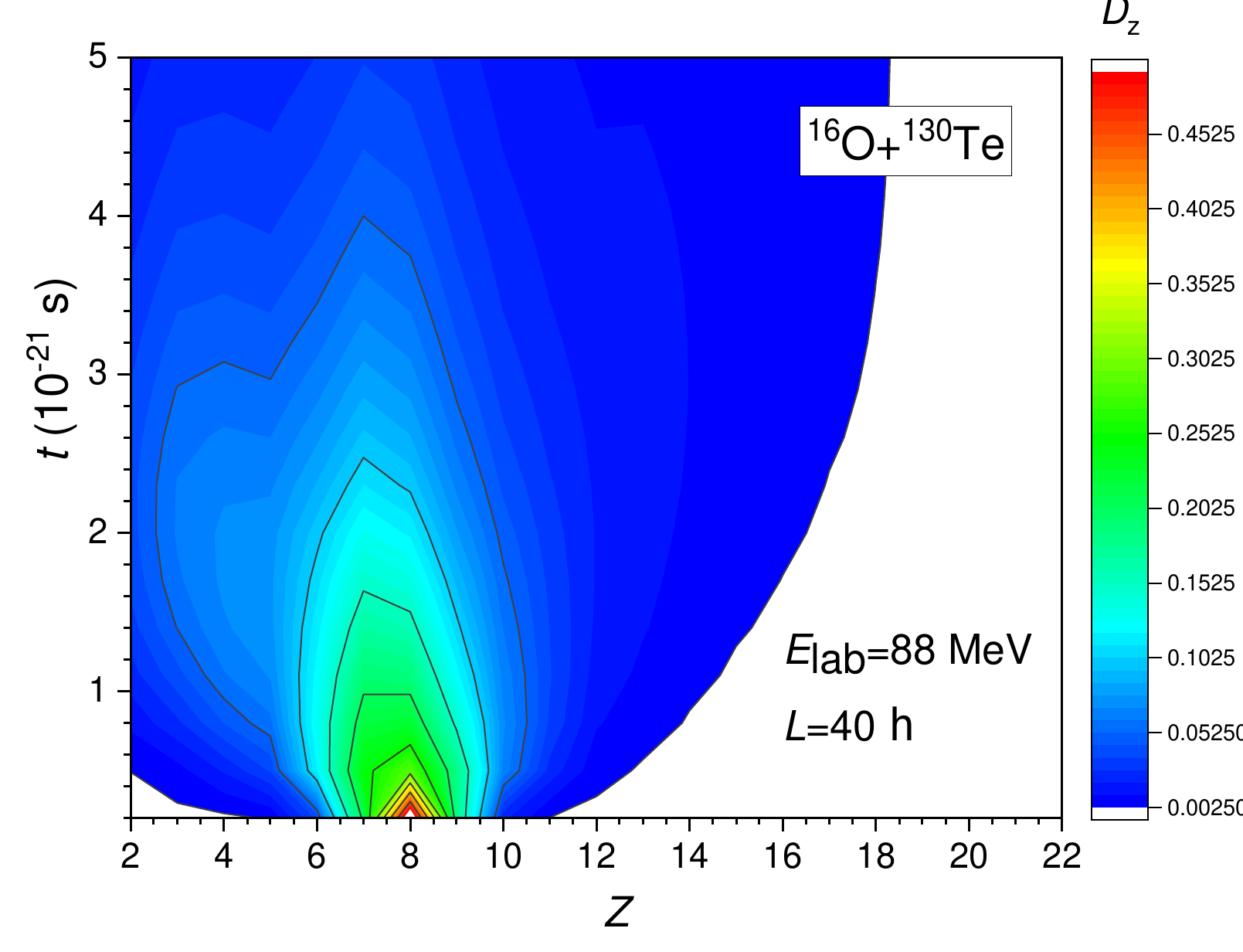}
\vspace{-0.2cm}
\caption{\label{DzL40E88} Evolution of the charge distribution for the
projectile-like fragments for the $^{16}$O+$^{130}$Te reaction at
$E_{\rm c.m.}=78.4$ MeV and $L=40 \hbar$. The results have been obtained for the orientation angles $\alpha_1=45^{o}$ and$\alpha_2=30^{o}$.
}
\end{figure}
The dependence of the barrier $B^*_{\rm fus}$ on $L$ is seen from the analysis
Fig. \ref{BqfBfus} which shows the PES values for the very
 asymmetric charge asymmetry ($Z \rightarrow 2$ and $Z_c \rightarrow 58$) increase
  strongly with $L$ due to smallness of the moment of inertia DNS with the
  $\alpha$ particle.
 Therefore, the fusion probability decreases by the increase of $L$,
 since the intrinsic fusion barrier $B^*_{\rm fus}$ increases
 and quasifission barrier $B_{\rm qf}$ decreases (see Fig. \ref{BqfBfus}) by
 increasing $L$.
 This circumstance is a reason leading to the creation of the
 favorable range of the angular momentum for the incomplete fusion
 in the  peripheral collisions  at the large beam energies.
  The enhance of the rotational energy in the PES with $L$
  is  related with the strongly decrease of moment of inertia
  of the very asymmetric  DNS:
  $J_{\rm DNS}(Z,A,Z_c,A_c)=\mu R_m^2+(J_1(Z,A)+J_2(Z_c,A_c))/2$, which
  is used in calculation  of the rotational energy:
 \begin{equation}
V_{\rm rot}(Z,A,Z_c,A_c,L,R)=
\frac{L(L+1)}{2J_{\rm DNS}(Z,A,Z_c,A_c)}.
  \end{equation}
  The reduced mass of DNS and moments inertia of the
  interacting nuclei are calculated by the expressions
  $\mu=m A A_c/(A+A_c)$, $J_1=1/5 m A (a_1^2+b_1^2)$ and $J_2=1/5 m A_c (a_2^2+b_2^2)$,
  respectively; $m$ is a nucleon mass; $a_i$ and $b_i$ are small and large
  radii of nuclei;   $R_m$ is the distance corresponding to the minimum of the
  potential well of the nucleus--nucleus interaction;
  $\alpha_i$ and $\beta_i$ are the orientation angle of the axial symmetry axis and
  the deformation parameter of the DNS fragments, respectively.

 The  excitation energy $E^{*}_{Z}$
 of DNS at the given value of the beam energy is  calculated taking into
 account the change in the intrinsic energy of DNS at the change of nucleon
 numbers of fragments:
\begin{eqnarray}
E^*_{Z}(E_{\rm c.m.},A,L,\{\beta_i,\alpha_i\})&=&E_{\rm c.m.}+\Delta Q_{\rm
gg}(Z,A)\nonumber\\
&-&V(Z,A,R_m,L,\{\beta_i,\alpha_i\}),\nonumber\\
\label{Edns}
\end{eqnarray}
where $\Delta Q_{\rm gg}(Z,A)=B+B_c-B_P-B_T$;
$B_P$, $B_T$, $B$ and $B_c$  are binding energies 	of the initial
($``P''$ and ``$T$'')  and interacting  fragments; $V(Z,A,R_m,L,\{\beta_i,\alpha_i\})$
 is the minimum value of the potential well, and it
  is a function of the nuclear shape $\beta_i$ and orientation angles
  $\alpha_i$ of the axial symmetry axis of the deformed nuclei to the beam
   direction  \cite{Nasirov2005}.
   The probability of the  $\alpha$-particle formation and its yield
   has been estimated as  a quasifission fragment  by the formula
\begin{eqnarray}\label{yield}
Y_{Z}^{}(E^*_{Z},A,L,t) &=&\Lambda^{\rm qf}_{Z}
(B^{\rm qf}(Z,A,\{\alpha_i,\beta_i\}),T_Z(A,\alpha_i,\beta_i))\nonumber\\
 &\times&\sum_{k=0}^{k_{\rm max}}D_{Z}(A,E^*_{Z},L,t_0+k \Delta t)
\end{eqnarray}
%
where $D_Z^{}(A,E^*_{Z},L,t)$ is the probability of population of the
DNS configuration $(Z, Z_{\rm CN}-Z)$ for a given set of $E^*_{Z}$ and $L$;
a value of $k_{\rm max}$ corresponds to the interaction time $t$ of the DNS
fragments when  $D_Z^{}(A,E^*_{Z},L,t_{\rm int})< 10^{-5}$, {\it i.e.}
the DNS has gone to complete fusion or it has broken up as quasifission products
 (see Fig. \ref{DzL40E88}). The part of $D_Z$ going to region $Z<2$ is
 a contribution to the complete fusion.
The evolution of the charge distribution $D_Z$ is calculated by the
transport master equation \cite{Nasirov2016} with initial conditions
$D_{Z}(A,E^*_{Z},L,t=0)=1$ for $Z=Z_P(Z_T)$ and $A=A_P(Z_T)$;
   $\Lambda^{\rm qf}_Z$  is the width of the decay through the quasifission barrier
    which is calculated by the expression
\begin{eqnarray}
 \Lambda_Z(B^{\rm qf}(Z,A,\{\alpha_i,\beta_i\}),T_Z(\alpha_i,\beta_i))\propto\nonumber\\
\exp\left(\frac{-B_{\rm qf}(Z,A,\{\alpha_i,\beta_i\})}
 {T_Z(\alpha_i,\beta_i)}\right),
\end{eqnarray}
where  $T_Z$ is the effective temperature of the DNS with the charge asymmetry $Z$:
$T_Z(A,\alpha_i,\beta_i) = 3.46 \sqrt{E^*_{Z}(A,\alpha_i,\beta_i)/A_{\rm tot}}$,\, $A_{\rm
tot}=A_P+A_T$. The transition coefficients 	of the transport master equation depend
on the energy, spin, and occupation numbers of the single-particle states of the
nucleons in the DNS fragments (see Refs. \cite{Nasirov2016,Kayumov2022} for details).
The occupation numbers of nucleons in the DNS fragments depend on $T_Z$.
\begin{figure}[t!]
\centering
  \includegraphics[width=0.47\textwidth]{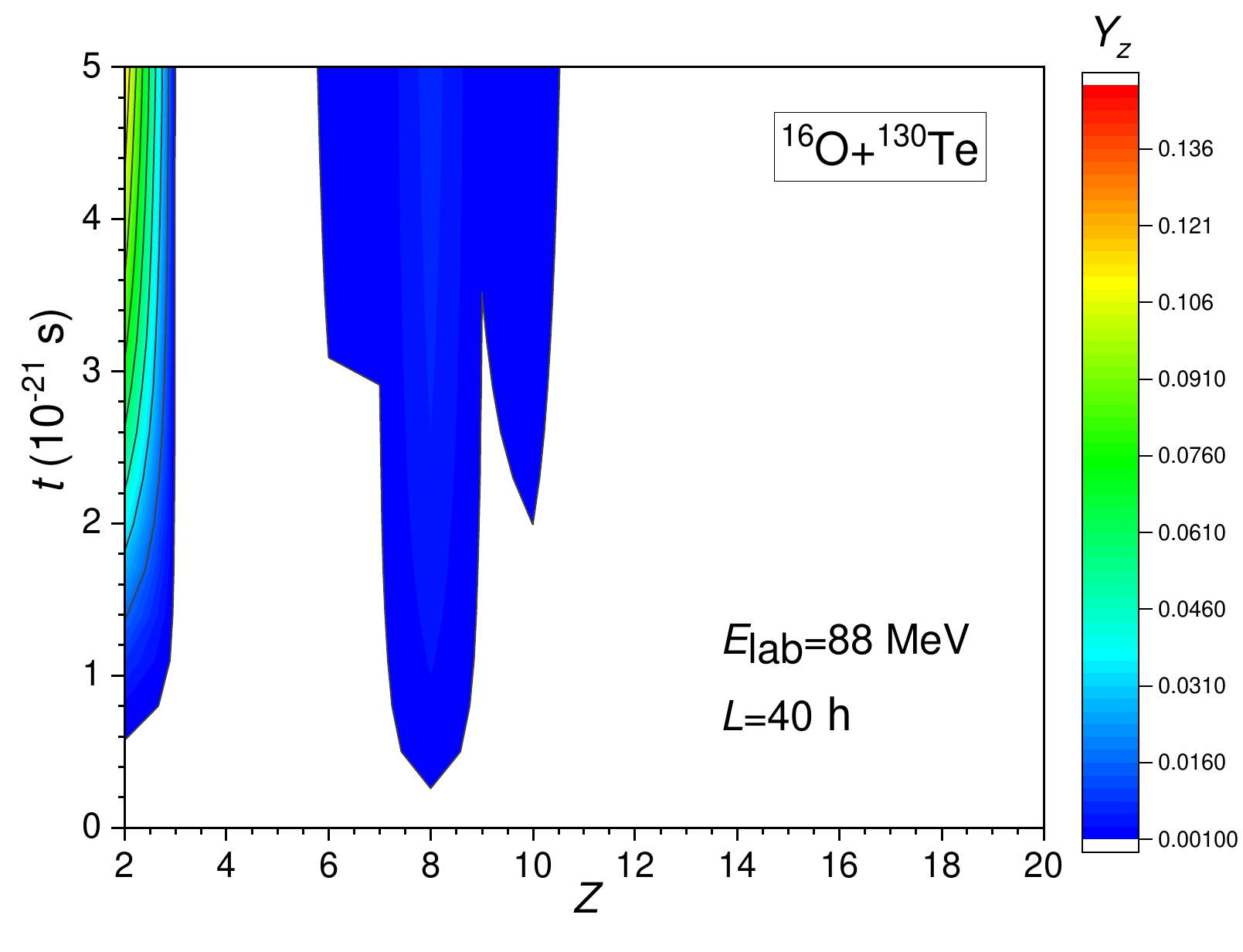}
\vspace{-0.2cm}
\caption{Evolution of the yield of $\alpha$ particle $Y_Z(Z=2)$ calculated for the
$^{16}$O+$^{130}$Te reaction at $E_{\rm c.m.}=78.4$ MeV and $L=40 \hbar$.
The results have been obtained for the orientation angles $\alpha_1=45^{o}$ and
$\alpha_2=30^{o}$.
\label{Yz16O130Te}}
\end{figure}

Evolution of the charge distribution $D_Z$ of the DNS and yield $Y_Z$ of
fragments calculated for the $^{16}$O+$^{130}$Te reaction
at $E_{\rm c.m.}=78.4$ MeV and $L=40 \hbar$ are presented in Fig. \ref{Yz16O130Te}.
The presented results have been obtained for the orientation angles $\alpha_1=45^{o}$ and
$\alpha_2=30^{o}$.

The knowledge of the $Y_Z(Z=2)$ values as
a function of $E_{\rm c.m.}$  and $L$ allows us to find the partial cross sections
of the incomplete fusion as a very mass asymmetric channel of the quasifission
process by the following expression:
\begin{eqnarray}\label{ParICF}
\sigma_{\rm ICF}(E_{\rm c.m.},L) =
\sigma_{\rm cap}(E_{\rm c.m.},L)*Y_Z (E_{\rm c.m.},L),
\end{eqnarray}
where $\sigma_{\rm cap}(E_{\rm lab},L)=\pi \lambda\hspace{-2.3 mm}^{-2}
 {\mathcal P}_{\rm cap}(E_{\rm lab},L)$, where
 $\lambda\hspace{-2.3 mm}^{-2}$ is the de Broglie wavelength
 corresponding to the collision energy $E_{\rm lab}$ and
  ${\mathcal P}_{\rm cap}$ is the capture probability
  which is found from the calculation of the collision trajectory
  for  the given values of $E_{\rm c.m.}$ and orbital angular
  momentum $L$~\cite{Kayumov2022}.

  The increase of the probability of the mass and charge
  distribution  at the $Z=2$ and $A=4$ corresponding to the $\alpha$ particle
  occurs in the collisions with $L=$40--60 $\hbar$ for the $^{16}$O+$^{130}$Te reaction
  and in the collisions with $L=$30--40 $\hbar$ for the $^{18}$O+$^{93}$Nb reaction.
  These results allow us to make theoretical analysis of the incomplete
  fusion mechanism by the calculation of the evaporation residues excitation
  function and to compare with the measured data from Refs. \cite{Singh2014,Agarwal}.
\begin{figure}[t!]
\centering
  \includegraphics[width=0.45\textwidth]{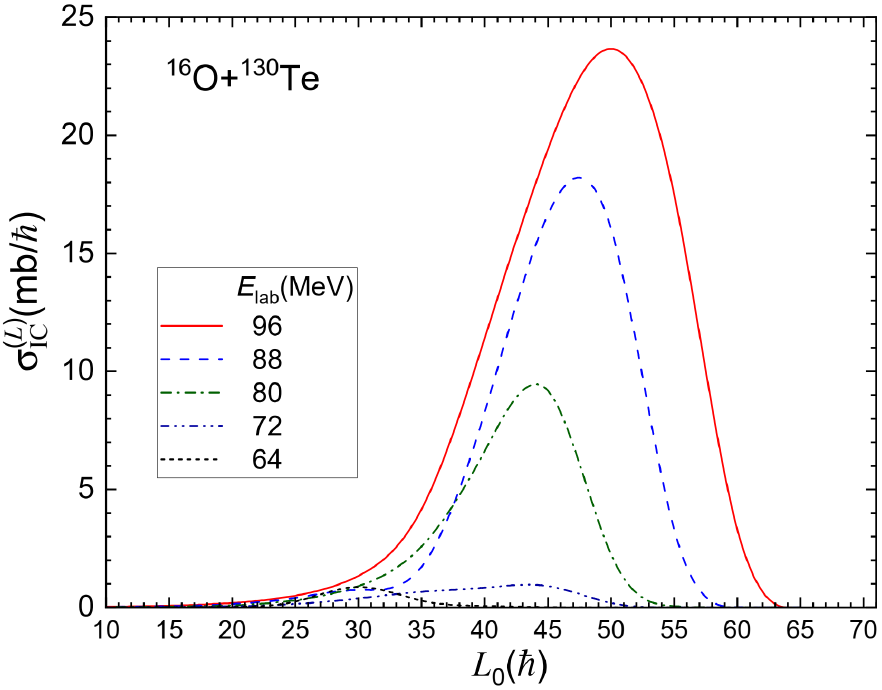}
  \vspace{-0.2cm}
\caption{The partial cross section of the incomplete fusion
$\sigma_{\rm ICF}(E_{\rm lab},L)$  as a function of the angular momentum $L$ for the set of
collision  energy values $E_{\rm lab}$ for the $^{16}$O+$^{130}$Te  reaction.
\label{ICPar16O130Te}}
\end{figure}

\begin{figure}[t!]
\centering
  \includegraphics[width=0.45\textwidth]{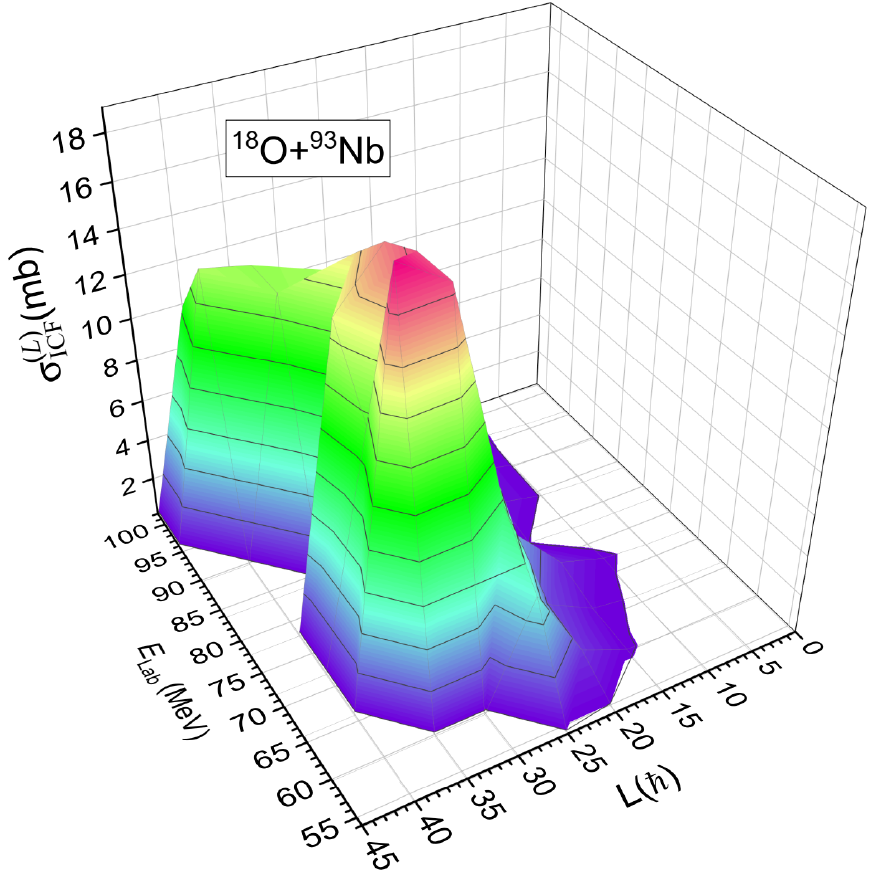}
  \vspace{-0.2cm}
\caption{The partial cross section of the incomplete fusion
$\sigma_{\rm ICF}(E_{\rm lab},L)$  as a function of the angular momentum $L$ for the set of
collision  energy values $E_{\rm lab}$ for the $^{18}$O+$^{93}$Nb  reaction.
\label{ParICF18O93Nb}}
\end{figure}

 The  partial cross sections of the incomplete fusion
  for the $^{16}$O+$^{130}$Te and $^{18}$O+$^{93}$Nb  reactions are presented in Figs.
   \ref{ICPar16O130Te} and \ref{ParICF18O93Nb}, respectively.
 The range of the angular momentum values $L=$35--60 $\hbar$
 and $L=$25--40 $\hbar$ are favorable  for the incomplete mechanism in the first
 and second reactions, respectively.
    These ranges are in agreement with the measured
    averaged values $<L>=40 \hbar$  and $<L>=30 \hbar$ in the $^{14}$N+$^{159}$Tb
    \cite{Inamura1977}  and  $^{16}$O+$^{146}$Nd  \cite{Barker} reactions.
    The authors of these last two experiments have measured
    $\gamma$-multiplicity to estimate the averaged value of the
    angular momentum corresponding to the incomplete fusion mechanism.

   The other important result are ranges of the  DNS excitation energy $E^*_Z$
   obtained from the calculations of the capture and fusion cross sections.
   \begin{figure}[t!]
\centering
  \includegraphics[width=0.47\textwidth]{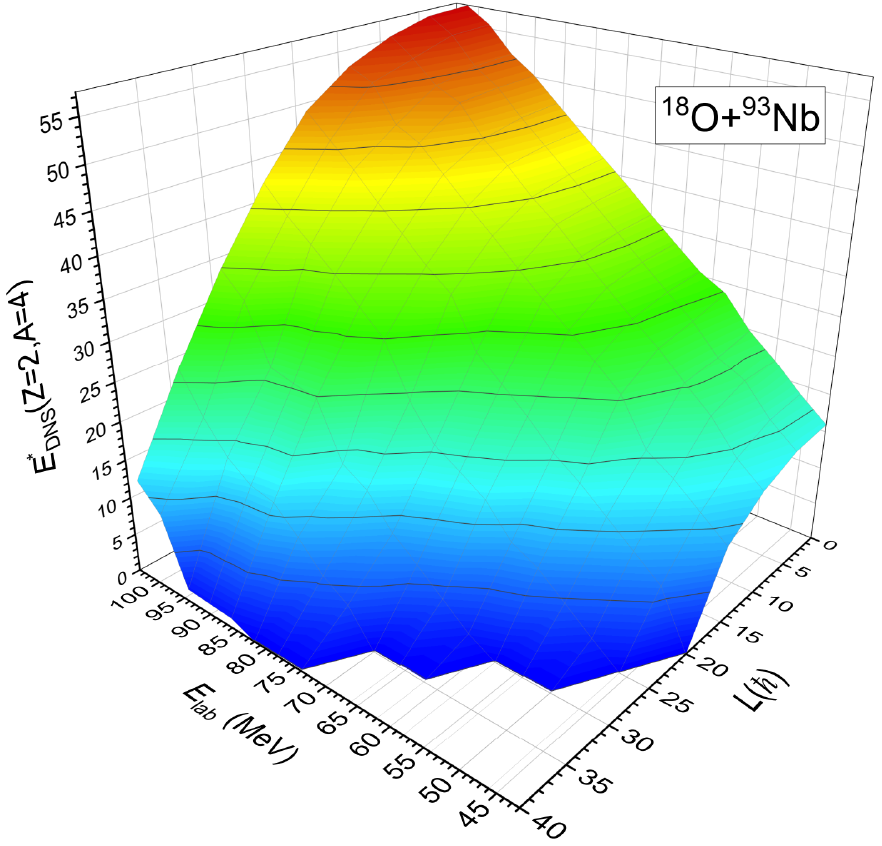}
  \vspace{-0.2cm}
\caption{The range
  $E^*_Z(E_{\rm lab},L,\{\beta_i,\alpha_i\})$=10--50 MeV of
  the DNS excitation energy  for the angular momentum
  $L=$35--50 $\hbar$  leading to the incomplete fusion in the $^{18}$O+$^{93}$Nb
  reaction by emission  of $\alpha$ particle from the interacting system.
 \label{Edns18O93Nb}}
\end{figure}

   The cross sections of the   evaporation residues of  the nuclei formed at the incomplete
   fusion
  are calculated  using the DNS excitation energy $E^*_Z$ which
    decreases due to the increase in the rotational energy of the
  interacting system (see Fig. \ref{Edns18O93Nb}). The important result is
   the formation of the wide range
  $E^*_Z(E_{\rm lab},\ell,\{\beta_i,\alpha_i\})$=10--50 MeV of
  the DNS excitation energy  for the angular momenta
  $L=$35--50 $\hbar$  of DNS corresponding to the incomplete fusion.
  The nearly plateau of the  excitation function of the evaporation
  residues of the 1n--5n channels for the wide of the beam energy
  (65--105 MeV) is related with this phenomenon.

The ER  cross sections of  the  $x$n channels of the incomplete fusion
accompanied with the emission of  the $\alpha$ particle  have been calculated
 in the framework of the DNS model \cite{Kayumov2022}.
 We should note that the excitation energy of the conjugate nucleus after emission
 of $\alpha$ particle is $E^*_{\rm ICF}(L)=E^*_{Z=2}(E_{\rm c.m.},L,\{\beta_i,\alpha_i\})$
 for the given values of orientation angles $\alpha_i$ of the DNS fragments.
 The evaporation residue (ER) cross section of the $x$n channel ($x$ neutrons have
 been emitted) has an excitation energy $E^*{(x)}$ and its value is
 calculated a sum of the partial cross sections:
\begin{eqnarray}\label{erpar}
\sigma^{(x)}_{\rm ER}(E^*_x)=\sum^{L_d}_{\ell=0}(2L+1)\sigma^{(x)}_{\rm ER}(^*_x,L),
\end{eqnarray}
where $\sigma^{(x)}_{\rm ER}(^*_x,L)$ is the partial cross section of the ER
formation as the survival cross section of the intermediate nucleus
at each step $x$ of the de-excitation cascade by the formula \cite{Kayumov2022,Mandaglio2012}
\begin{eqnarray}\label{ercs}
\sigma^{(x)}_{\rm ER}(E^*_{x},L)=\sigma^{x-1}_{\rm ER}(E^*_{x-1},L)W^{(x)}{\rm
sur}(E^*_{x-1},L).
\end{eqnarray}
Here, $\sigma^{x-1}_{\rm ER}(E^*_{x-1},L)$ is the partial cross section of the intermediate excited nucleus formation at the $(x-1)$th step, and  $W^{(x)}_{\rm sur}(E^*_{x-1},L)$ the survival probability of the $x$th intermediate nucleus against fission along each step of the de-excitation cascade. It is calculated by the statistical model implanted in KEWPIE2 \cite{Kewpie2}.

Obviously $\sigma^{(0)}_{\rm ER}(E^*_{\rm ICF},L)=\sigma_{\rm ICF}(E^*_{\rm ICF},L)$
which is calculated by (\ref{ParICF}).
   This procedure is similar to the calculation of
   the cross section of the evaporation residues formed after emission
of neutrons from the heated and rotating compound nucleus formed at complete fusion
\cite{Kayumov2022,Mandaglio2012} where the equation  $\sigma^{(0)}_{\rm ER}(E^*_{\rm CN},L)=\sigma_{\rm fus}(E^*_{CN},L)$ was used.
\begin{figure}[t!]
\centering
  \includegraphics[width=0.45\textwidth]{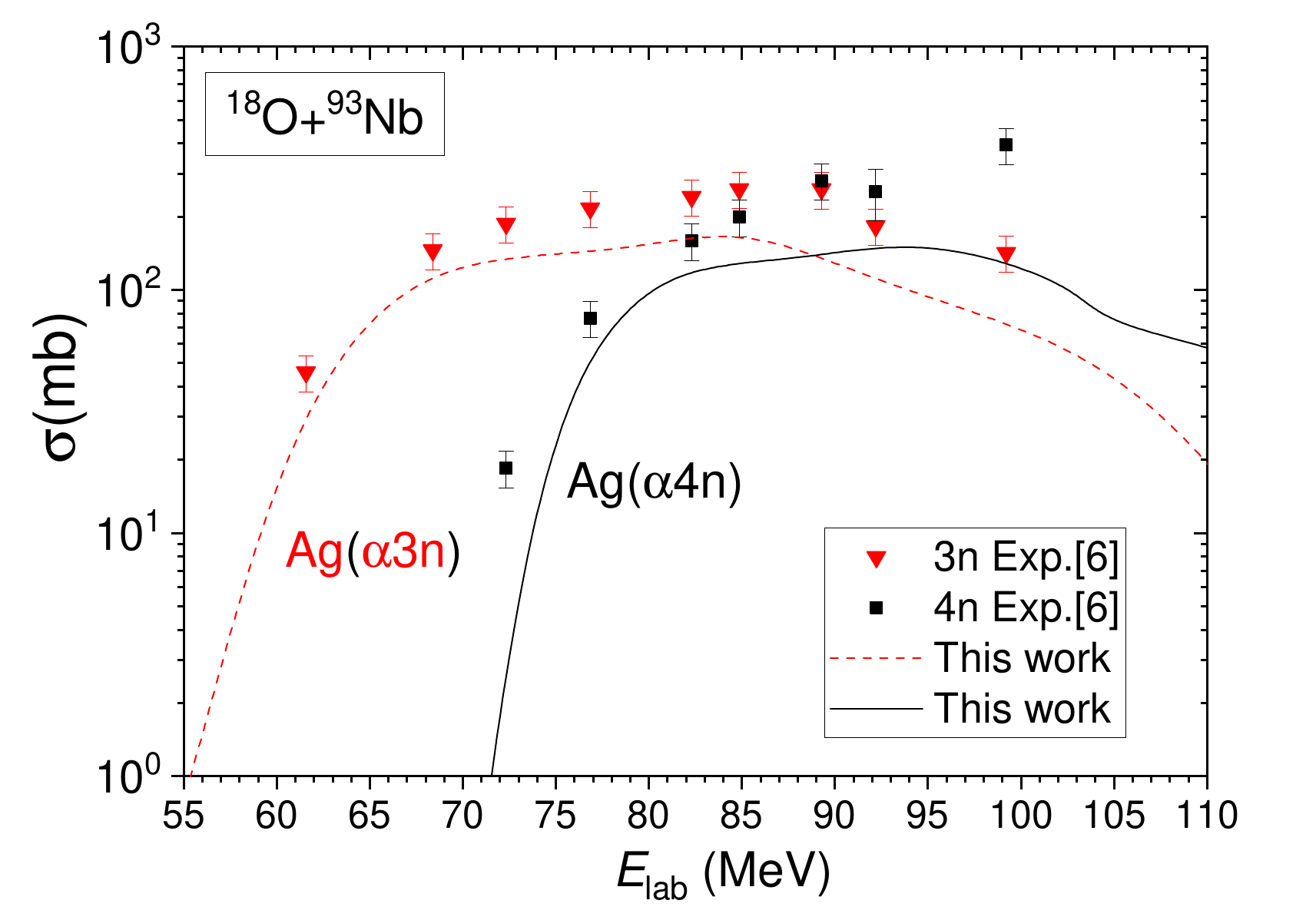}
  \vspace{-0.2cm}
\caption{Comparison of the theoretical cross sections (solid curve)
of the evaporation residues formed in the $^{18}$O+$^{93}$Nb
 incomplete fusion  reaction after emission of 3 neutrons with the
measured experimental data (squares) presented in Ref. \cite{Agarwal}.
\label{ERalpaXn}}
\end{figure}

In Fig. \ref{ERalpaXn}, the results of calculations using Eq. (\ref{erpar}) are compared
with  the measured cross sections of the evaporation residues of $^{108}$Ag and $^{107}$Ag
formed in the $^{18}$O+$^{93}$Nb incomplete fusion reaction after emission of 3 and 4
neutrons \cite{Agarwal}, respectively.
  The  agreement our results with the measured data is  better than that calculated in Ref.
  \cite{Agarwal} by the standard  methods ALICE-91 \cite{Kumar2019} and  PACE4
  \cite{Kumar2013,Agarwal}.
\begin{figure}[t!]
\centering
  \includegraphics[width=0.45\textwidth]{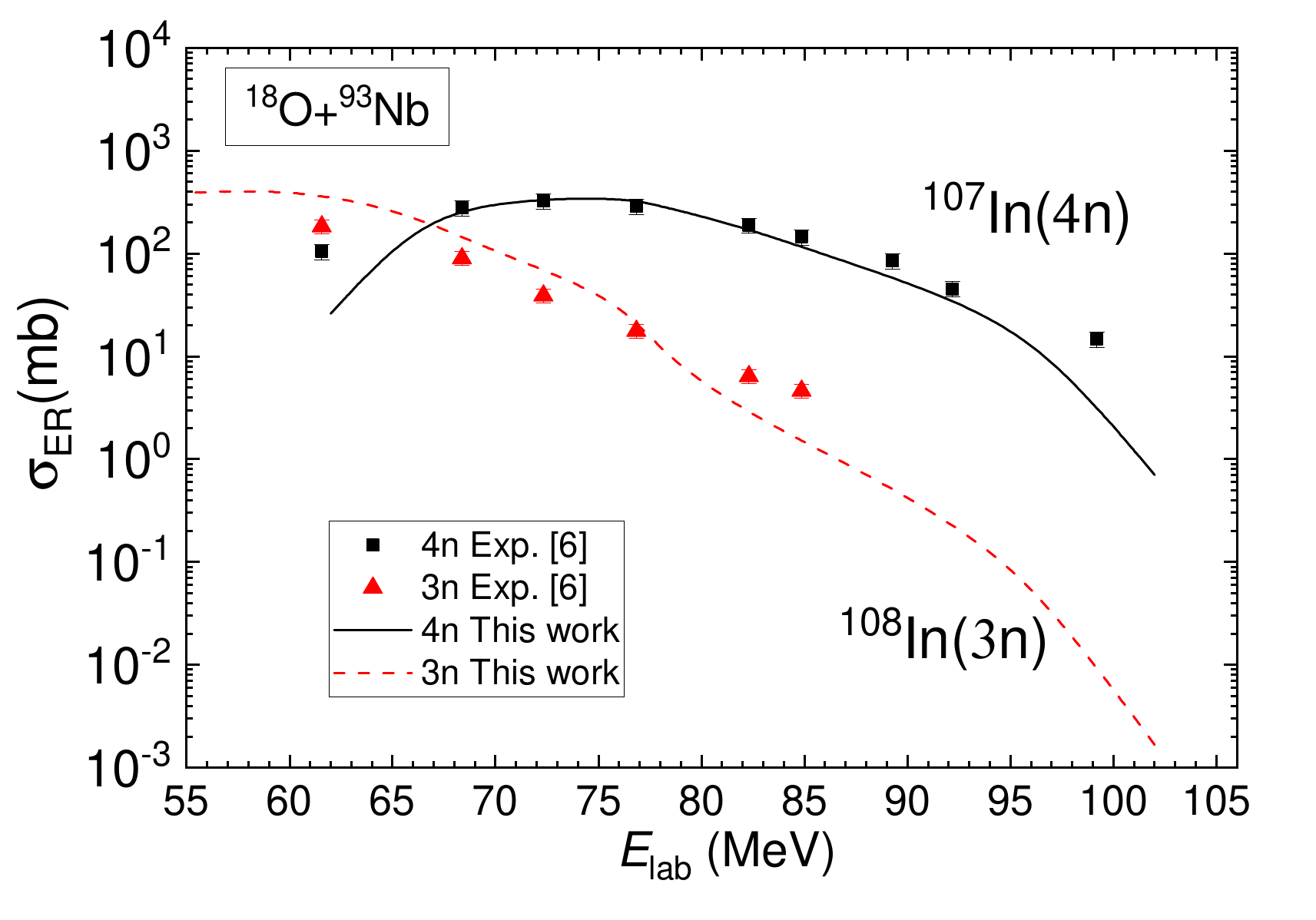}
  \vspace{-0.2cm}
\caption{Comparison of the theoretical cross sections (dashed and solid curves) of the
evaporation residues formed in the $^{18}$O+$^{93}$Nb  complete fusion
reaction after emission of 3 (triangles) and 4 (squares) neutrons with the
measured experimental data  presented in Ref. \cite{Agarwal}.
\label{CN34nER}}
\end{figure}
\begin{figure}[t!]
\centering
  \includegraphics[width=0.45\textwidth]{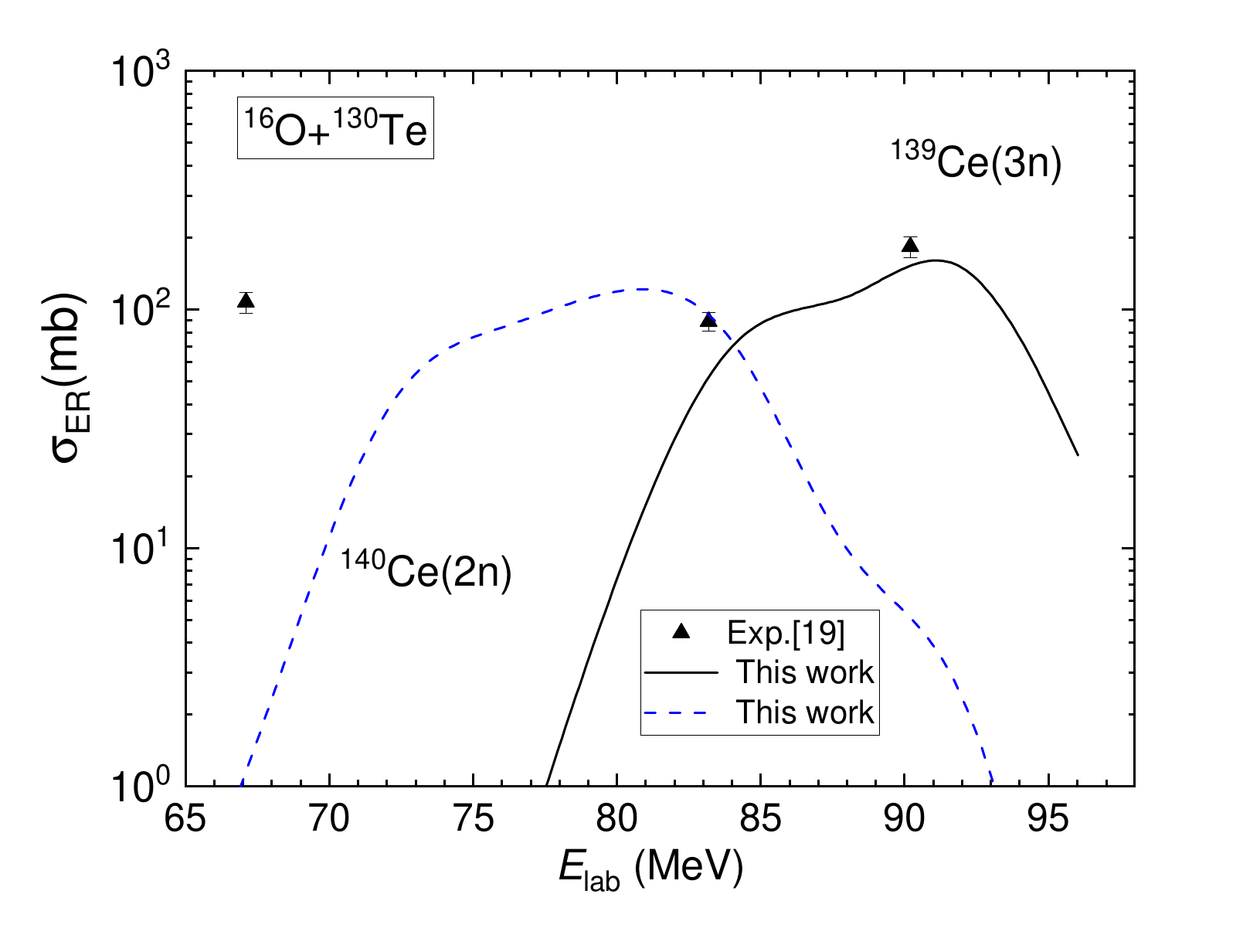}
  \vspace{-0.2cm}
\caption{Comparison of the theoretical cross sections (solid curve) of the evaporation
residues formed in the $^{16}$O+$^{130}$Te  complete fusion
reaction after emission of 3 neutrons with the
measured experimental data (triangles) presented in Ref. \cite{Singh2014}.
\label{ER3nCe}}
\end{figure}

In Fig. \ref{CN34nER}, the cross sections of the evaporation residues of  $^{107}$In and
$^{108}$In formed  in the $^{18}$O+$^{93}$Nb complete fusion reaction after emission of 4
and 3 neutrons,  respectively, are compared with the measured experimental data.
 It is seen from Fig. \ref{CN34nER} that the behaviours of the experimental data and
 theoretical curves do not have  a  plateau  as a function of the beam energy.
  From this point of view, the behaviours of the excitation functions
  of the evaporation residues formed in the incomplete and complete
  fusion reactions are different. The difference is explained
  by increasing the rotational energy $V_{\rm rot}(Z,L)$ of the DNS
  with the very mass-asymmetric configuration corresponding to
  the $\alpha$-particle emission in collisions with the large beam energies:
   $V_{\rm rot}(Z)$ takes an appreciable part of the
   kinetic energy of the relative motion. As a result, the
  residual nucleus  formed after emission of $\alpha$-particle in the incomplete
   fusion is less excited than compound nucleus formed in the complete fusion
    in heavy-ion collision with the same values of the orbital angular momentum
    and beam energy.
  The fission barrier $B_f$ of the heated and rotating compound nucleus
   decreases due to its large excitation  energy $E^*_{\rm CN}$ and angular momentum $L$.

  The theoretical excitation function of the ER
  $^{139}$Ce formed in the incomplete fusion of the $^{16}$O+$^{130}$Te
  reaction after emission of the $\alpha$ particle  and  3 neutrons
    is compared with the measured experimental data \cite{Singh2014} in Fig. \ref{ER3nCe}.
    Our approach does not allow us to reach an agreement at low energies
    where the excitation function of the ER $^{140}$Ce formed
   after emission of the $\alpha$ particle  and  2 neutrons dominates over 3n neutron
   channel.

  The new mechanism of the incomplete fusion has been explored
  as a very mass-asymmetric quasifission of a DNS formed
 at the capture of the projectile nucleus (full momentum transfer)
   by the target nucleus.
     The $L$-dependence of the charge distribution of the DNS
  fragments leads to formation of its configuration consisting of the  $\alpha$ particle
  and a conjugate nucleus.
 The centrifugal force related with the rotation of  a very mass-asymmetric DNS
 is strong for the $L>30$. Consequently, it leads to the incomplete fusion
 which can be considered  as the quasifission producing $\alpha$ particle and
 a  conjugate heavy fragment.
 This phenomenon is related with the transformation  of a significant part of the
 kinetic energy of the collision energy to the rotational energy of the DNS
 formed at capture of the projectile by target nucleus.
 As a result the conjugate fragment is less heated than the compound nucleus
 formed in the  complete fusion.
 Therefore, the plateau of the  excitation function of the ER
 of the 3n--4n channels of the incomplete fusion in the high energy  range
  is observed for the $^{18}$O+$^{93}$Nb and $^{16}$O+$^{130}$Te reactions.
  The  measured cross sections of ER formed in the incomplete fusion and complete
 fusion channels have been reproduced well by the DNS model and the statistical model
  implanted in KEWPIE2 \cite{Kewpie2}.

  Our exploration of the incomplete mechanism has confirmed
  a decisive role of the orbital angular momentum in the reaction mechanisms
  of the heavy-ion collisions at energies above the Coulomb barrier and below 10 MeV/nucleon.

\bibliographystyle{apsrev4-2}
\bibliography{References_PRL}
\end{document}